\documentclass{iau}
\usepackage{graphicx}

\def\lb {$L_{\rm bol}$}

\title[Black hole masses from X-rays] %% give here short title %%
{Black hole masses from X-rays}

\author[Xin-Lin Zhou \& Roberto Soria]  %% give here short author list %%
{Xin-Lin Zhou $^1$
\and  Roberto Soria$^2$}

\affiliation{$^1$ National Astronomical Observatories, Chinese
Academy of Sciences, \\ Beijing 100012, China  \\ email: {\tt zhouxl@nao.cas.cn} \\[\affilskip]
$^2$International Centre for Radio Astronomy Research, Curtin
University \\GPO Box U1987, Perth, WA 6845, Australia\\email: {\tt
roberto.soria@icrar.org}}

\pubyear{2012}
\volume{290}  %% insert here IAU Symposium No.
\jname{Feeding compact objects: Accretion on all scales}
\editors{C.M. Zhang, T. Belloni, M. M\'endez \& S.N. Zhang, eds.}

\begin{document}

\maketitle

\begin{abstract}
We discuss two methods to estimate black hole (BH) masses using X-ray
data only: from the X-ray variability amplitude and from the photon
index $\Gamma$.
%together with $L_{\rm 2-10keV}$.
The first method is based
on the anti-correlation between BH mass and X-ray variability amplitude.
Using a sample of AGN with BH masses from reverberation mapping,
we show that this method shows small intrinsic scatter.
%, no larger than the
%uncertainties of black hole masses used.
The second method is based on the correlation between $\Gamma$ and
both the Eddington ratio $L_{\rm bol}/L_{\rm Edd}$ and the bolometric
correction \lb/$L_{\rm 2-10keV}$.
%, where \lb~ is the
%bolometric luminosity and $L_{\rm 2-10keV}$ is the 2-10 keV
%luminosty) and Eddington ratio.
%(Zhou \& Zhao 2010).
%Then the BH mass
%can be estimated from \ga~ together with $L_{\rm 2-10keV}$.

\end{abstract}

\firstsection % if your document starts with a section,
%\section{Introduction}
%We introduce two methods to estimate black hole (BH) mass using the
%X-ray data only. The first method is based on the correlation
%between the BH mass and X-ray variability amplitude (XVA, so-called
%excess variance, Nandra et al. 1997). It was found that this
%correlation for the reverberation-mapped (RM) active galactic nuclei
%(AGNs) shows small intrinsic scatter (Zhou et al. 2010; Ponti et al.
%2012); suggesting that XVA is a good mass indicator. The second
%method is based on the correlation between photon index (\ga) and
%both bolometric correction (\lb/$L_{\rm 2-10keV}$, where \lb~ is the
%bolometric luminosity and $L_{\rm 2-10keV}$ is the 2-10 keV
%luminosty) and Eddington ratio (Zhou \& Zhao 2010). Then the BH mass
%can be estimated from \ga~ together with $L_{\rm 2-10keV}$.
\section{X-ray variability amplitude}
\noindent
The XVA $\sigma^2_{\rm rms}$ (also known as ``excess variance'')
is the variance of a light curve normalized by its mean squared
after correcting for experimental noise (Nandra et al. 1997;
Turner et al. 1999). For a light-curve segment with $N$ bins:
\begin{equation}
\sigma^2_{\rm rms} = \frac{1}{N\mu^2}\,\sum_{i=1}^{N}
\left[\left(X_i - \mu\right)^2 - \sigma_i^2\right],
\end{equation}
where $X_i$ and $\sigma_i$ are count rates and uncertainties in each
bin, and $\mu$ is the average of the count rates. There is an
empirical (anti)correlation between XVA and BH masses in AGN
(\cite{luyu2001,ONeill2005}). To constrain this correlation, we
selected and studied two (largely overlapping) AGN samples with
X-ray observations longer than 40 ks: one sample with BH masses
derived from reverberation mapping, and the other from the $M_{\rm
BH}$-$\sigma_{\ast}$ relation (see \cite{Zhou10a} for details). We
found that the intrinsic dispersion of the $M_{\rm
BH}$-$\sigma^2_{\rm rms}$ relation for the reverberation-mapped AGN
is quite small, no larger than the uncertainties in the BH masses:
\begin{equation}
M_{\rm BH} = 10^{4.97\pm0.26}\, \left(\sigma^2_{\rm rms}\right)^{-1.00\pm0.10}\, M_{\odot}.
\end{equation}
A similar result was independently obtained
by \cite[Ponti et al. (2012)]{Ponti2012}.

We used this relation to determine the BH mass in the Seyfert
galaxies MCG-6-30-15 and 1H0707$-$495, using archival {\it
XMM-Newton} data. We obtained BH masses of $(2.6\pm0.5) \times 10^6
M_{\odot}$ and $(6.8\pm 0.7) \times 10^5 M_{\odot}$, respectively
(\cite{Zhou10a}). The XVA derived from multiple {\it XMM-Newton}
observations changes by a factor of 2--3. This means that the
uncertainty in the BH mass from a single observation is slightly
worse than that from the reverberation mapping or the stellar
velocity dispersion methods. However, if the XVA randomly scatters
around the true value for the power spectral density (PSD), 
the mean XVA of many data segments reduces the error 
(\cite{Vaughan2003}). We conclude that the XVA might be a better BH mass
estimator than the virial method.

The $M_{\rm BH}$-$\sigma^2_{\rm rms}$ relation is explained by a
shift of the high-frequency break $f_{\rm b}$ in the PSD, to lower
frequencies for higher BH masses. $f_{\rm b}$ scales approximately
as $\dot{m}/M_{\rm BH}$, where $\dot{m}$ is the dimensionless
Eddington accretion rate (\cite{McHardy2006}); however, we found
that there is no or very weak correlation between XVA and $\dot{m}$,
confirming the findings of \cite[O'Neill et al. (2005)]{ONeill2005}.
This suggests that the normalization of the PSD varies with
$\dot{m}$ in a way that compensates for the break-frequency
dependence on $\dot{m}$.

Finally, we point out that our sample of AGN is skewed towards BHs
with low mass and high Eddington rates. In forthcoming work, we will
explore: where the relation saturates at the low-mass end
(\cite{Ai2011}); how it may extend to very-low-luminosity nuclear
BHs in normal galaxies; and whether it may be used to estimate BH
masses in ultraluminous X-ray sources (ULXs).

%\begin{figure}[a]
% \vspace*{-2.0 cm}
%\begin{center}
% \includegraphics[width=2.5in, angle=270]{f1.ps}
% \vspace*{-1.0 cm}
% \caption{Correlation between the XVA and the BH mass. The filled points denote the 21
%AGNs with the BH masses measured from the reverberation-mapping
%method. The intrinsic dispersion of the best-fit linear relation for
%the 21 AGNs given by the Nukers' estimate is $\sim$0.2 dex. We also
%plot the objects with the mass estimated from the empirical virial
%relation(open circles).}
%   \label{fig1}
%\end{center}
%\end{figure}

%\begin{figure}[b]
% \vspace*{-2.0 cm}
%\begin{center}
% \includegraphics[width=2.in, angle=270]{f2.ps}
% \vspace*{-1.0 cm}
% \caption{The distribution of XVA derived
%from multiple XMM observations for the Seyfert galaxy MCG-6-30-15
%(solid line) and 1H0707-495 (dotted line), corresponding to a BH
%mass of $(2.1\pm1.0) \times 10^6 M_{\odot}$ and $(7.1\pm 2.2) \times
%10^5 M_{\odot}$, respectively.}
%   \label{fig2}
%\end{center}
%\end{figure}

\vspace{-0.4cm}

\section{BH mass estimates from $\Gamma$ and $L_{\rm 2-10keV}$}
\noindent
X-ray spectral studies of accreting BHs show
a correlation between the photon index $\Gamma$ of the power-law
component
%(generally measured over the $2$--$10$ keV range)
and the Eddington ratio.
%(either $L_{\rm 2-10keV}/L_{\rm Edd}$ or $L_{\rm bol}/L_{\rm Edd}$).
Specifically, $\Gamma$ varies from
$\approx 2.5$ for $L_{\rm bol} \sim L_{\rm Edd}$ to $\approx 1.5$
for $L_{\rm bol} \sim 10^{-2} L_{\rm Edd}$ (for AGN: \cite{Shemmer2008,GuCao2009};
for stellar-mass BHs: \cite{WuGu2008}). (At even lower luminosities,
there is evidence that $\Gamma$ increases again,
but with a much weaker correlation: \cite{GuCao2009,Corbel2008}).
In \cite{Zhou10b}, we refined this correlation
by choosing a sample of 29 low-redshift ($z < 0.33$) AGN in the luminosity range
$10^{-2} \sim L_{\rm bol}/L_{\rm Edd} \sim 1$. We determined
$\Gamma$ by fitting their {\it XMM-Newton}/EPIC spectra
in the $2$--$10$ keV range. We selected only radio-quiet AGN,
because beaming in radio-loud sources may affect measurements of
the intrinsic value of $\Gamma$. All sources have BH masses from
reverberation mapping, and $L_{\rm bol}$ estimated from simultaneous
X-ray, UV and optical data. We obtain:
\begin{equation}
\log\left(L_{\rm bol}/L_{\rm Edd}\right) = (2.09 \pm 0.58)\,\Gamma
- (4.98 \pm 1.04)
\end{equation}\\[-30pt]
\begin{equation}
\log\left(L_{\rm bol}/L_{\rm 2-10keV}\right) = (1.12 \pm
0.30)\,\Gamma - (0.63 \pm 0.53).
%\log\left(L_{\rm bol}/L_{\rm 2-10keV}\right) = (2.52 \pm 0.08)\,\Gamma
%- (3.12 \pm 0.15)
\end{equation}

%Zhou \& Zhao (2010) suggested that \ga~ acts as an indicator of
%bolometric correction, \lb/$L_{\rm 2-10keV}$, in radio-quiet AGNs.
%Correlations between $\Gamma_{\rm 2-10keV}$ and both \lb/$L_{\rm
%2-10keV}$ and Eddington ratio are presented, based on simultaneous
%X-ray, UV and optical observations of RM AGNs.
The correlation (2.2) appears stronger than (2.1) (see also
\cite{Jin2012}). In forthcoming work, we shall compare these
correlations with those inferred from high-redshift AGN, to check
for evolutionary effects in AGN spectral properties. Assuming no
evolutionary effect, we can use (2.1,2.2) to determine BH masses in
AGN for which we know the X-ray luminosity. We estimate a mean
uncertainty in the BH mass of a factor of 2 or 3
(\cite{Shemmer2008}). We shall also explore the $\Gamma$ versus
$L_{\rm bol}/L_{\rm Edd}$ correlation for $L_{\rm bol}/L_{\rm Edd}
\gtrsim 1$, with possible applications to ULXs and quasars.

\end{document}